\documentstyle[aps,prl,epsf,floats,twocolumn]{revtex}
\begin{document}
\draft
\twocolumn[\hsize\textwidth\columnwidth\hsize\csname @twocolumnfalse\endcsname
\title{Disorder-driven non-Fermi liquid behavior in Kondo alloys
}
\author{E. Miranda, V. Dobrosavljevi\'{c}}
\address{National High Magnetic Field Laboratory, Florida State
University\\1800 E. Paul Dirac Dr., Tallahassee, Florida 32306.}
\author{G. Kotliar}
\address{Serin Physics Laboratory,
Rutgers University,
PO Box 849,
Piscataway NJ, 08855.
}
\date{\today}
\maketitle

\begin{abstract}

We demonstrate that a model of disordered Anderson lattices can account
for many non-Fermi liquid features observed in a number of Kondo alloys.
Due to the exponential nature of the Kondo temperature scale, even
moderate disorder leads to a rather broad distribution of Kondo
temperatures, inducing strong effective disorder seen by the conduction
electrons.  Spins with very small Kondo temperatures remain unquenched
and dominate the low temperature properties.  The model predicts
logarithmic divergences in thermodynamic quantities at low temperatures.
We also find a {\em linear} temperature dependence of the resistivity, a
feature that remained a stumbling block in previous theoretical
attempts.  We argue that for realistic amounts of disorder, such
marginal Fermi liquid behavior is a very robust feature of disordered
Kondo alloys.

\end{abstract}

]

Non-Fermi liquid (NFL) behavior in metals represents one of the key
unresolved issues in condensed matter physics. There exists by now a
large class of non-magnetic metallic f-electron materials which do not
behave as Fermi liquids at low
temperatures\cite{nfl_ucupd,bernal,nfl_mupd3,nfl_lacecu2si2,nfl_cethrhsb,%
nfl_cecuau,nfl_uthpd2al3,nfl_cepdsi}. In some of them, the proximity to
a $T=0$ quantum critical point appears to be the origin of the anomalous
behavior\cite{nfl_cecuau,nfl_cepdsi,qcpth}. However, in several other
cases, NFL behavior only occurs when the system has been sufficiently
alloyed so that it is not close to any phase boundary.  This is the case
of the alloys UCu$_{5-x}$Pd$_x$\cite{nfl_ucupd,bernal},
M$_{1-x}$U$_x$Pd$_3$ (M = Sc,Y)\cite{nfl_mupd3,nfl_uthpd2al3},
La$_{1-x}$Ce$_x$Cu$_{2.2}$Si$_2$\cite{nfl_lacecu2si2},
Ce$_{1-x}$Th$_x$RhSb\cite{nfl_cethrhsb}
U$_{1-x}$Th$_x$Pd$_2$Al$_3$\cite{nfl_uthpd2al3}.  In all of these
systems the specific heat varies as $C(T)/T \approx a {\rm ln}(T_0/T)$
and the resistivity is linear with a large zero temperature intercept
$\rho(T) \approx \rho_0 (1 - T/T_0)$. The magnetic susceptibility has
been often fitted by a logarithm or a weak power law.

Some attempts have been made to explain the anomalous low-temperature
properties based on exotic one-impurity mechanisms, such as the
quadrupolar Kondo model\cite{quad}. Inconsistencies with the predictions
of the model for the resistivity ($\approx \sqrt{T}$) and in an applied
magnetic field in some of these systems, however, invite the
consideration of other mechanisms for NFL behavior.

Quite generally, the {\em large residual resistivity} of these systems
together with their alloy nature immediately suggest that disorder could
be significant.  In an important recent study\cite{bernal}, the strong
broadening of the copper NMR line of UCu$_{5-x}$Pd$_x$ ($x = 1$ and
$1.5$) has provided an independent indication of the essential role
played by disorder in at least one of these compounds. These results
suggested the presence of strong spatial fluctuations in the
characteristic Kondo temperature $T_K$ of the local
moments\cite{TKdist}. Indeed, by using a model distribution function
$P(T_K)$ and well-known single-impurity results, they were then able to
quantitatively describe the low temperature thermodynamic properties
(specific heat and magnetic susceptibility) as well as the NMR
linewidths.  The proposed picture implicitly assumes independent local
moments which is usually sufficient for understanding the thermodynamics
of most heavy fermion compounds. Of course, in the context of transport
in concentrated Kondo systems, such an assumption appears to be
unjustified, since it cannot be reconciled with the well established
coherence effects at low temperatures.

The central question addressed in this letter is whether disorder
effects can explain not only the thermodynamics, but also {\em the
anomalous transport} in these systems.  We will formulate a theory
appropriate for {\em concentrated} magnetic impurities, which can
describe the coherence effects in the clean limit.  We will show that
correlation effects will strongly enhance any extrinsic disorder,
generating an extremely broad distribution of Kondo temperatures.  This
leads to the destruction of coherence and, for sufficient disorder, to
the breakdown of Fermi liquid behavior. The low temperature properties
can be viewed as resulting from a {\em dilute gas} of localized
elementary excitations: those Kondo spins that remain unquenched. This
picture of dirty Kondo lattices, similar in spirit to the original
Landau description of simple metals, provides a clear theoretical
underpinning for one route towards marginal Fermi liquid behavior.

We start with a disordered non-degenerate infinite-U Anderson lattice
model 
\begin{eqnarray}
H &=& \sum\limits_{\bbox{k}\sigma} \epsilon({\bbox{k}})
c^{\dagger}_{\bbox{k}\sigma} c^{\phantom{{\dagger}}}_{\bbox{k}\sigma}
+ \sum\limits_{j\sigma} E^f_j f^{\dagger}_{j\sigma}
f^{\phantom{{\dagger}}}_{j\sigma} \nonumber \\
&+& \sum\limits_{j\sigma} V_j (c^{\dagger}_{j\sigma}
f^{\phantom{{\dagger}}}_{j\sigma}  + {\rm H. c.} ),
\label{hammy}
\end{eqnarray}
where, $c_{\bbox{k}\sigma}$ destroys a conduction electron with momentum
$\bbox{k}$ and spin $\sigma$ from a broad uncorrelated band with
dispersion $\epsilon({\bbox{k}})$ and half bandwidth $D$ and
$f_{j\sigma}$ destroys an f-electron at site $j$ with spin $\sigma$. The
infinite-U constraint at each f-orbital is assumed ($n^f_j \le 1$).  The
on-site energies $E^f_j$ and the hybridization matrix elements $V_j$ are
assumed to be distributed according to some distribution functions
$P_1(E_f)$ and $P_2(V)$. In the Kondo limit, the local Kondo temperature
is given by $ T_{Kj} = D \exp(E^f_j/(2\rho_0 V^2_j))$ ($\rho_0 \approx
\frac{1}{2D}$) and will be correspondingly distributed. Because of the
strong scattering off the f-sites, disorder in f-parameters is dominant
and we will thus neglect other types of disorder in the c-band.

To analyze the properties of our model, we focus on the dynamical
self-consistent theory of strong correlation and
disorder\cite{lisa_disorder+interaction}. The problem can then be
reduced to an ensemble of one-impurity problems in a self-consistently
generated self-averaging bath of conduction electrons.  The equations
simplify considerably in the case of a semi-circular conduction density
of states, where the ensemble of impurity problems is governed by the
action
\begin{eqnarray}
S^{\rm imp}_j &=& \sum_{\omega_n} \left[
f^{\dagger}_{j\sigma}
\left( - 
i\omega_n + E^f_j + \Delta_j(i\omega_n) \right)
f^{\phantom{{\dagger}}}_{j\sigma}
\right], 
\label{effaction}
\end{eqnarray}
where the infinite-U constraint is implied and
\begin{eqnarray}
\Delta_j(\omega) &=& \frac{V^2_j}{\omega - t^2
\overline{G}_c(\omega)}.
\label{delta}
\end{eqnarray}
Here, $t$ is the hopping parameter and $\overline{G}_c(\omega)$ is the
disorder-averaged local conduction electron Green's function. The latter
is determined self-consistently by
\begin{eqnarray}
\overline{G}_c(\omega) &=& \left\langle \frac{1}{\omega - t^2
\overline{G}_c(\omega) - \Phi_j(\omega)} \right\rangle^{\rm av},
\label{gc}
\end{eqnarray}
where
\begin{eqnarray}
\Phi_j(\omega) &=&
\frac{ V^2_j}{\omega - E^f_j - \Sigma^{\rm imp}_{fj}(\omega)}.
\label{phi}
\end{eqnarray}
Here $\left\langle \dots \right\rangle^{\rm av}$ denotes the average
over disorder and $\Sigma^{\rm imp}_{fj}(\omega)$ is the f-electron
self-energy derived from the impurity model of Eq.~(\ref{effaction}).
In absence of disorder, these equations reduce to the dynamical
mean-field theory of the Anderson lattice\cite{lisa_interaction}, while
for $U=0$ they are equivalent to the CPA treatment of disorder
\cite{CPA} for the {\em conduction electrons}\cite{cpa-validity}.  In
general, the theory is exact in the limit of large coordination.  Once
$\overline{G}_c(\omega)$ has been determined, the conduction electron
self-energy $\Sigma_c(\omega)$ can be obtained from
\begin{eqnarray}
\overline{G}_c(\omega) = \int d\epsilon
\frac{\rho_0(\epsilon)}{\omega - \epsilon - \Sigma_c(\omega)},
\label{sigmac}
\end{eqnarray}
where $\rho_0(\epsilon) = \sqrt{1 - (\epsilon/2t)^2}/\pi t$.

Let us analyze the qualitative behavior of $\Phi_j(\omega)$.  From the
Fermi liquid analysis of the impurity problem, it is well known that
$\Sigma^{\rm imp}_{fj}(\omega=0)$ is a real quantity at
$T=0$\cite{hewson}. Therefore, one can write
\begin{eqnarray}
\Phi_j(\omega=0)&=& -\frac{V^2_j}{E^f_j+{\rm Re}[\Sigma^{\rm
imp}_{fj}(0)]}.
\label{FLphi}
\end{eqnarray}
$\Phi_j(0)$ measures the scattering strength at site $j$ at the Fermi
level. In the clean limit, $\Phi_j(\omega)$ will be the same at every
site and Eqs.~(\ref{gc}) and (\ref{sigmac}) give
$\Sigma_c(\omega)=\Phi(\omega)$.  In this case, $\Sigma_c(\omega=0)$ is
a real quantity, reflecting the coherent nature of the DC transport at
zero temperature.

\begin{figure}
\epsfxsize=3.2in \epsfbox{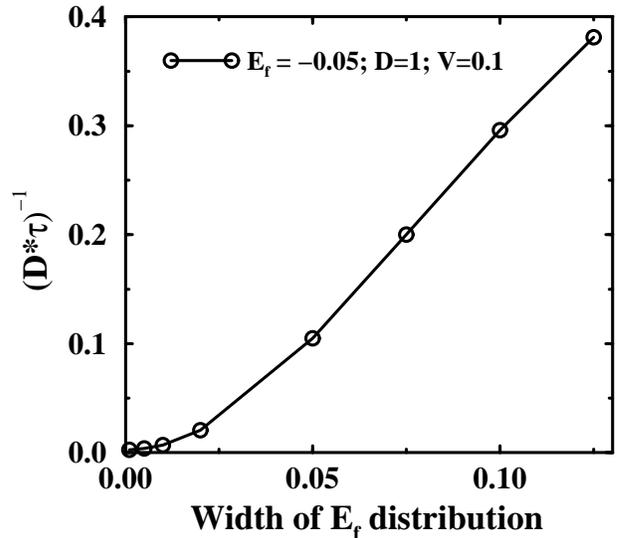}
\caption{Scattering rate as a function of the width of the $E_f$
distribution. The parameters used are shown in the figure. The strong
correlations in the f-shell produce an enhanced effective disorder.
\label{fig1}}
\end{figure}

By contrast, when the system is disordered, a distribution of scattering
strengths $\Phi_j$ is generated, strongly affecting the transport
properties. By applying the large-N mean field theory to the impurity
problems at zero temperature, we have solved the self-consistent problem
defined by Eqs.~(\ref{effaction}--\ref{phi}). The resulting scattering
rates as a function of the width of the $E_f$ distribution (for a fixed
uniform value of $V$) are shown in Fig.~\ref{fig1}.  Similar results are
obtained for a distribution of $V$ values holding $E_f$ fixed. For the
residual resistivities reported for the NFL alloys, e.  g.
UCu$_{5-x}$Pd$_x$\cite{nfl_ucupd}, one can estimate $D\tau
\approx 3-5$. Due to the strong f-shell correlations, rather large
scattering rates can be generated by a small disorder strength in
f-parameters (see Fig.~\ref{fig1} and Eq.~\ref{FLphi}).  Comparable
amounts of disorder, in the absence of correlations, cannot produce
these large resistivities.

Thus, with sufficient disorder, scattering off the f-sites becomes
incoherent and the resistivity assumes a monotically decreasing
dependence on temperature, resembling the single-impurity result. The
actual scattering rate, however, requires the solution of the full set
of Eqs.~(\ref{effaction}--\ref{phi}).  If $P(T_K)$ is broad enough, the
low-temperature dependence can be non-trivial. To analyze that, it is
useful to rewrite the above equations in terms of the impurity T-matrix
$T^{\rm imp}_j(\omega) \equiv V^2_j G_{fj}(\omega)$, where
$G_{fj}(\omega)$ is the f-Green's function computed from the action in
Eq.~(\ref{effaction}). We find
\begin{eqnarray}
\overline{G}_c(\omega) &=& \frac{1}{\omega - t^2 \overline{G}_c(\omega)}
+ \frac{\langle T^{\rm imp}_j(\omega)\rangle^{\rm av}}{(\omega - t^2
\overline{G}_c(\omega))^2}
\label{gc_t}
\end{eqnarray}
and, from Eq.~(\ref{sigmac}),
\begin{eqnarray}
\Sigma_c(\omega) &=& \frac{\langle T^{\rm imp}_j(\omega)\rangle^{\rm
av}}{\overline{G}_c(\omega)(\omega - t^2 \overline{G}_c(\omega))}.
\label{sigmac3}
\end{eqnarray}

We now raise the temperature slightly from 0 to $T$ and denote
corresponding variations by $\delta_T$.  Then
\begin{mathletters}\label{dsigdgc}
\begin{eqnarray}
\delta_T\Sigma_c(\omega) &=&
\left.\frac{1-t^2G_c^2(\omega)}{G_c^2(\omega)}\right|_{T=0} \delta_T
G_c(\omega); \label{dsig}\\
A(\omega)\delta_T G_c(\omega)& -& \int d\omega' B(\omega,\omega') \delta_T
G_c(\omega') \nonumber \\ &=& 
\left. \langle \delta_T T^{\rm imp}_j(\omega) \rangle^{\rm av}
\right|_{G_c^0},
\label{dgc}
\end{eqnarray}
\end{mathletters}
where 
\begin{mathletters}\label{aandb}
\begin{eqnarray}
A(\omega) &=& \bigg\{ t^2+\left[\omega-t^2G_c(\omega)\right]
\left[\omega-3t^2G_c(\omega)\right] \nonumber \\ &-&
\left. \textstyle
\frac{t^2\langle\left[T^{\rm imp}_j(\omega)\right]^2\rangle^{\rm av}}
{\left[\omega-t^2G_c(\omega)\right]^2} \bigg\} \right|_{T=0};
\label{aom}\\ 
B(\omega,\omega')&=& \textstyle \left.\langle \frac{\left[T^{\rm
imp}_j(\omega)\right]^2}{V^2_j} 
\frac{\delta \Sigma^{\rm imp}_{fj}(\omega)}
{\delta G_c(\omega')}
\rangle^{\rm av}
\right|_{T=0}.
\label{bom}
\end{eqnarray}
\end{mathletters}
Here, the temperature dependence of the self-energy is expressed in
terms of the temperature dependence of the disorder-averaged T-matrix.
In general, the self-consistency condition couples different
frequencies, as seen in the integral equation~(\ref{dgc}).  However, the
{\em leading low temperature behavior} is determined only by the $\omega
=0$ component of the averaged T-matrix, so in the following we
concentrate on this object.

\begin{figure}
\epsfxsize=3.2in \epsfbox{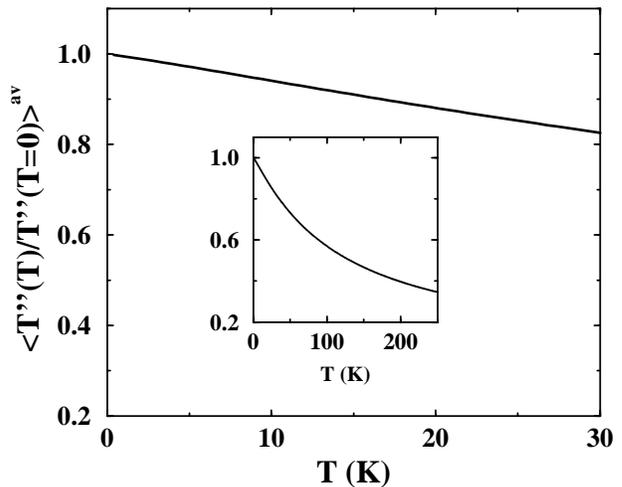}
\caption{Temperature dependence of the imaginary part of the single
impurity T-matrix averaged over the disorder distribution appropriate
for UCu$_{3.5}$Pd$_{1.5}$, as determined experimentally in
Ref.~\protect\cite{bernal}. The inset shows the same quantity over a
wider temperature range.
\label{fig2}}
\end{figure}

Fig.~\ref{fig2} shows the result of averaging the imaginary
part the single impurity T-matrix over the distribution of Kondo
temperatures deduced from the experiments of Ref.~\cite{bernal}. For the
single impurity dependence, we used a simple scaling form with the
correct asymptotic behavior at high and low temperatures. The dependence
is linear at low temperatures.

It is easy to understand the origin of the linear behavior and why it
does not depend on the detailed shape of the curve. We will focus on the
imaginary part of the impurity T-matrix since it gives the dominant
contribution. It has the following scaling form
\begin{equation}
T_{\rm imp}''(T) = \frac{\sin^2\delta_0}{\pi\rho_0} t(\frac{T}{T_K}),
\label{scalingT-mat}
\end{equation}
where $\delta_0$ is the phase shift at $T=0$.  The function $t(x)$ has
the following asymptotics
\begin{eqnarray}
t(x) &\approx& \left\{
\begin{array}{ll}
1 - \alpha x^2 & \qquad x \ll 1 \\
\frac{\beta}{({\rm ln}(x))^2}  & \qquad x \gg 1\\
\end{array} \right. ,
\label{asymp}
\end{eqnarray}
where $\alpha$ and $\beta$ are universal numbers. It follows that
\begin{equation}
\delta_T T_{\rm imp}'' = - \frac{\sin^2\delta_0}{\pi\rho_0} \left[1 - 
t(\frac{T}{T_K})\right] \equiv - \frac{\sin^2\delta_0}{\pi\rho_0}
F(T/T_K) 
\end{equation}

\begin{figure}
\epsfxsize=3.2in \epsfbox{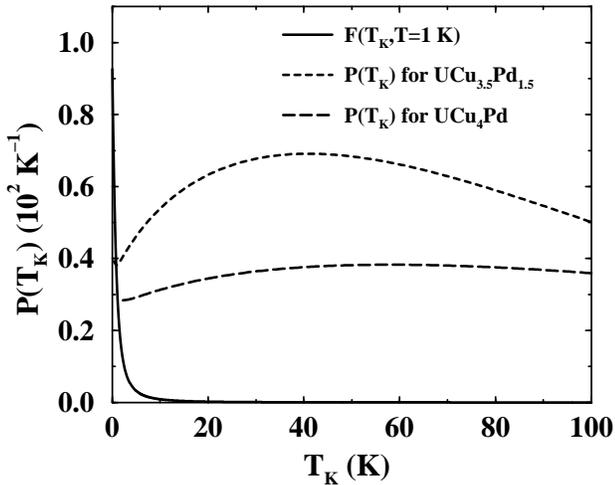}
\caption{Comparison of the experimentally determined distribution of
Kondo temperatures of the alloys UCu$_{5-x}$Pd$_x$ ($x=1,1.5$) (from
Ref.~\protect\cite{bernal}) with the function $F(T_K,T)$ defined in the
text. The function $F(T_K,T)$ only probes the $T_K=0$ value of the
distributions at low $T$.
\label{fig3}}
\end{figure}

Now, {\em for a fixed temperature T and as a function of $T_K$}
\begin{eqnarray}
F(T/T_K) &\approx& \left\{
\begin{array}{ll}
\frac{\alpha T^2}{T_K^2} & \qquad T_K \gg T \\
1 - \frac{\beta}{({\rm ln}(T/T_K))^2}  & \qquad T_K \ll T
\end{array} \right. .
\label{Fasymp}
\end{eqnarray}
$F(T/T_K)$ is strongly peaked at $T_K \approx 0$, decays as $1/T_K^2$
and has a width of order $T$ (see Fig.~\ref{fig3}). For low $T$, it can
be written in terms of a delta function of $T_K$, hence
\begin{equation}
\delta_T T_{\rm imp}'' \approx - \frac{a \sin^2\delta_0}{\pi\rho_0} T
\delta(T_K),
\label{Fasdelta}
\end{equation}
where $a = \int dx F(1/x) $. When inserted into Eq.~(\ref{dsig}) this
yields 
\begin{equation}
\delta_T \Sigma_c \approx -\frac{i a P(0)}{\pi\rho_0A_0} T.
\label{final}
\end{equation}

Therefore, the low temperature dependence only probes the $P(T_K)$
distribution at low values of $T_K$ as is clear from Fig.~\ref{fig3}.
In that region, $P(T_K)$ can be taken to be a constant and the
temperature can be scaled out of the average, yielding the negative
linear term.  As long as the distribution of Kondo temperatures is wide
enough so that $P(0)$ is appreciable, there will be a sizable linear
range. For sufficiently weak disorder, $P(0)$ is zero or negligible and
Fermi liquid behavior is recovered.

Physically, it is clear what is happening.  As the temperature is
raised, {\em a few diluted} spins with $T_K < T$ are unquenched and
cease to contribute to the resistivity. The linear term essentially
counts the number of liberated spins. Since this number is small at low
temperatures, they form a dilute system of removed scatterers, whose
effect is additive (or rather, subtractive), rendering our treatment of
disorder essentially exact.  Thus, even though the zero temperature
resistivity is a functional of the whole distribution $P(T_K)$, the low
temperature linear behavior is a much more robust feature which depends
only on the low $T_K$ tail $P(0)$.

Within the dynamical mean field theory, it is possible to show that the
picture of independent f-sites of Ref.~\cite{bernal} is justified for
thermodyamic quantities. An argument similar to the ones above then
gives $\chi(T) \sim {\rm ln} (T_0/T)$ and $C_V(T)/T \sim {\rm ln}
(T_0/T)$. Again in this case, the NFL behavior is due to the presence of
very low-$T_K$ spins.

It is important to comment on the potential limitations of this
approach.  In particular, we note that the dynamical mean field approach
cannot describe the effects of the RKKY interactions.  One could imagine
that pairs of local moments with very low $T_K$ could well condense into
RKKY singlets, affecting the temperature dependence.  However, if
$P(T_K)$ is very broad, the fraction of low-$T_K$ spins is very small
and they will be, in general, very far apart, rendering the RKKY
interaction less effective.

In summary, we have investigated the effects of disorder in concentrated
Kondo alloys, by solving dynamical mean field equations of disordered
correlated electrons. We find that disorder effects can lead to
considerable modifications of low temperature properties, leading to the
breakdown of conventional Fermi liquid behavior, consistent with some
Kondo alloys. Essentially, disorder fluctuations lead to the absence of
a characteristic energy scale associated with a conventional Fermi
liquid.

We acknowledge useful discussions with B. Andraka, N. Bonesteel, A. H.
Castro Neto and J. R. Schrieffer. This work was supported by the
National High Magnetic Field Laboratory at Florida State University.
GK was supported by NSF DMR 95-29138.

\end{document}